\documentclass[letterpaper,11pt]{article}

\usepackage{latexsym}
\usepackage{amsmath}
\usepackage{amssymb}
\usepackage{amsfonts}
\usepackage{hyperref}
\usepackage{xspace}
\usepackage{vmargin}
\usepackage[noend]{algorithmic}

\newtheorem{theorem}{Theorem}[section]
\newtheorem{lemma}[theorem]{Lemma}
\newtheorem{definition}[theorem]{Definition}
\newtheorem{observation}[theorem]{Observation}

\newenvironment{proof}[1]{%
  \begin{trivlist}{}{\setlength{\topsep}{0cm}\setlength{\partopsep}{0cm}}
  \item \textbf{#1.\@}\hspace*{1ex}\ignorespaces}%
  {\makebox[0cm]{}\nolinebreak\hfill$\Box$\end{trivlist}}

\newcommand{\ie}{i.e.,\xspace}
\newcommand{\eg}{e.g.,\xspace}

\newcommand{\ket}[1]{|{#1}\rangle} 
\newcommand{\tensor}{\ensuremath{\otimes}}

\newcommand{\norm}[1]{\ensuremath{\| #1 \|}}
\newcommand{\E}{\ensuremath{\operatorname{E}}}

\setmarginsrb{2cm}{2cm}{2cm}{3cm}{0cm}{0cm}{1cm}{1cm}

\title{Quantum Property Testing\thanks{Research done while all authors
were visiting the NEC Research Institute.} }
\addtocounter{footnote}{2}
\author{Harry Buhrman\thanks{CWI and University of Amsterdam; partially supported by the EU fifth framework project QAIP, IST-1999-11234.} \and
  Lance Fortnow\thanks{NEC Research Institute} \and 
Ilan Newman\thanks{Haifa University and NEC Research Institute} \and 
\addtocounter{footnote}{-3}
Hein R\"ohrig\footnotemark}

\date{November 12, 2001}

\begin{document}

\maketitle

\begin{abstract}
A language $L$ has a property tester if there exists a probabilistic
algorithm that given an input $x$ only asks a small number of bits
of $x$ and distinguishes the cases as to whether $x$ is in $L$ and $x$
has large Hamming distance from all $y$ in $L$. We define a similar
notion of quantum property testing and show that there exist languages
with quantum property testers but no good classical testers. We also show
there exist languages which require a large number of queries even for
quantumly testing.
\end{abstract}

\section{Introduction}
Suppose we have a large data set, for example, a large chunk of the
world-wide web or a genomic sequence. We would like to test whether
the data has a certain property, but we may not have the time to even
look at the entire data set or even a large portion of it.

To handle these types of problems, Rubinfeld and
Sudan~\cite{Rubinfeld-Sudan} and Goldreich, Goldwasser and
Ron~\cite{GGR} have developed the notion of property testing. Testable
properties come in many varieties including graph properties
(\eg~\cite{GGR,AFKS,F2,FN,anb,oded3}), algebraic properties of
functions~\cite{BLR,Rubinfeld-Sudan,EKKRV} and regular
languages~\cite{AIKS}. Ron~\cite{Ron} gives a nice survey of this
area.

In this model, the property tester has random access to the $n$ input
bits similar to the black-box oracle model. The tester can query only
a small, usually some fixed constant, probabilistically-chosen set of bits of
the input. Clearly we cannot determine from these small number of bits
whether the input sits in some language $L$. However, for many
languages we can distinguish the cases that the input is in $L$ from
whether the input differs from all inputs in $L$ of the same length by
some constant fraction of input bits.

Since we have seen many examples where quantum computation gives us an
advantage over classical
computation~\cite{BV,Simon-quantum,Shor,Grover} one may naturally ask
whether using quantum computation may lead to better property testers.
By using the quantum oracle-query model developed by Beals, Buhrman,
Cleve, Mosca and de~Wolf~\cite{BBCMW}, we can easily extend the
definitions of property testing to the quantum setting.

Beals, et.\ al.~\cite{BBCMW} have shown that for all total functions
we have a polynomial relationship between the number of queries required
by quantum machine and that needed by a deterministic machine. For greater
separations one needs to require a promise in the input and the known
examples, such as those due to Simon~\cite{Simon-quantum} and
Bernstein-Vazirani~\cite{BV}, require considerable structure in the
promise. In property testing there is a natural promise of either being in the
language or far from any input in the language. This promise would
seem to have too little structure to give a separation but in fact we
can prove that quantum property testing can greatly improve on classical
testing.

We show that any subset of Hadamard codes has a quantum property
tester and most subsets would require $\Theta(\log n)$ queries to test
with a probabilistic tester. This shows that indeed quantum property
testers are more powerful than classical testers. Moreover, we also
give an example of a language where the quantum tester is
exponentially more efficient. 

Beals, et.\ al.~\cite{BBCMW} observed that any $k$-query quantum
algorithm gives rise to a degree-$2k$ polynomial in the input bits,
which gives the acceptance probability of the algorithm; thus, a
quantum property tester for $P$ gives rise to a polynomial that is on
all binary inputs between $0$ and $1$, that is at least $2/3$ on
inputs with the property $P$ and at most $1/3$ on inputs far from
having the property $P$. Szegedy~\cite{szegedy:testingpolyconj}
suggested to algebraically characterize the complexity of classical
testing by the minimum degree of such polynomials; however, our
separation results imply that there are for example properties, for
which such polynomials have constant degree, but for which the best
classical tester needs $\Omega(\log n)$ queries.

Perhaps every language has a quantum property tester with a small
number of queries. We show that this is not the case. We prove that
for most properties of a certain size, any quantum algorithm requires
$\Omega(n)$ queries. We then show that a natural property, namely, the
range of a $d$-wise independent pseudorandom generator cannot be
quantumly tested with less than $(d+1)/2$ queries for any odd $d \le n
/ \log n - 1$.

\section{Preliminaries}

Property testing was first developed by Rubinfeld and
Sudan~\cite{Rubinfeld-Sudan} and Goldreich, Goldwasser and
Ron~\cite{GGR}.  We will use the following formal definition of
property testing from Goldreich~\cite{Goldreich-testing}.
\begin{definition}\label{propdef}
  Let $S$ be a finite set, and $P$ a subset of functions mapping $S$
  to $\{ 0, 1 \}$. A \emph{property tester} for $P$ is a
  probabilistic oracle machine $M$, which given a distance parameter
  $\epsilon > 0$ and oracle access to an arbitrary function $f : S
  \rightarrow \{ 0, 1 \}$, satisfies the following conditions:
  \begin{enumerate}
  \item the tester accepts $f$ if it is in $P$:
    \[
    \text{if } f \in P \text{ then } \Pr ( M^f (\epsilon) = 1 ) \ge
    \frac{2}{3}
    \]
  \item the tester rejects $f$ if it is far from $P$:
    \[
    \text{if } |\{ x \in S : f(x) \ne g(x) \} | > \epsilon \cdot | S |
    \text{, for every } g \in P \text{, then } \Pr ( M^f (\epsilon) =
    1 ) \le \frac{1}{3}
    \]
  \end{enumerate}
\end{definition}
\begin{definition}\label{propcompldef}
  The complexity of the tester is the number of oracle queries it
  makes: A property $P$ has a \emph{$(\epsilon,q)$-tester} if there is
  a tester for $P$ that makes at most $q$ oracle queries for
  distance parameter $\epsilon$.
  
  We often consider a language $L \subseteq \{ 0, 1 \}^*$ as the
  family of properties $\{ P_n \}$ with $P_n$ the characteristic
  functions of the length-$n$ strings from $L$, and analyze the query
  complexity $q=q(\epsilon, n)$ asymptotically for large $n$.
\end{definition}
To define quantum property testing we simply modify
Definition~\ref{propdef} by allowing $M$ to be a quantum oracle
machine. We need to be careful to make sure our oracle queries are
unitary operations. If $|f(x)|=|g(y)|$ for all $x, y \in S$ and $f, g
\in P$, we use the oracle-query model by Beals, Buhrman, Cleve, Mosca
and de~Wolf~\cite{BBCMW}: we define the unitary transformation $O_f$
that maps the basis state $\ket{x, y, z}$ to $\ket{x, y \oplus f(x),
  z}$ where $|x|=\lceil \log |S| \rceil$, $|y|=|f(x)|$ and $\oplus$
denotes bitwise exclusive or. In case there are $x, y, f, g$ so that
$|f(x)| \ne |g(y)|$, we define $O_f$ as mapping $\ket{x, l, y, z}$ to
$\ket{x, l + |f(x)| \mod k, y \oplus 0^{k-|f(x)|}f(x), z}$ where $k =
\max \{ |f(x)| : f \in P \text{ and } x \in S \}$, $|x| = \lceil \log
|S| \rceil$, $|l| = \lceil \log k \rceil$, and $|y| = k$.

We recommend the book of Nielsen and Chuang~\cite{NC} for background
information on quantum computing and the surveys of Ron~\cite{Ron} and
Goldreich~\cite{Goldreich-testing} for a background on property
testing.

\section{Separating Quantum and Classical Property Testing}

We show that there exist languages with $(\epsilon,
\operatorname{O}(1))$ quantum property testers that do not have
$(\epsilon, \operatorname{O}(1))$ classical testers.

\begin{theorem}
\label{septhm}
There is a languages $L$  that is $\epsilon$-testable by a quantum  test
with $O(1/\epsilon)$ number of queries but for which any
probabilistic $1/3$-test requires $\Omega(\log n)$ queries.
\end{theorem}
We use Hadamard codes to provide examples for Theorem~\ref{septhm}:
\begin{definition}
The \emph{Hadamard code} of $y \in \{ 0, 1 \}^{\log n}$ is $x = h(y) \in \{
0, 1 \}^n$ such that $x_i = y \cdot i$ where $y \cdot i$ denotes the
inner product of two vectors $y, i \in \mathbb{F}_2^{\log n}$.
\end{definition}
Note: the Hadamard mapping $h:\{0,1\}^{\log n} \rightarrow
\{0,1\}^n$ is one-to-one.
Bernstein and Vazirani~\cite{BV} showed that a quantum computer can
extract $y$ with one query to an oracle for (the bits of) $x$, whereas
a classical probabilistic procedure needs $\Omega(\log n)$ queries.
Based on this separation for a decision problem we construct for $A
\subseteq \{ 0, 1 \}^{\log n}$ the property $P_A \subseteq \{0,1\}^n$,
\[
P_A := \{ x : \exists y \in A \text{ s.t. } x=h(y) \} .
\]
Theorem~\ref{septhm} follows from the following two lemmas.
\begin{lemma}
\label{qylem}
For every $A$, $P_A$ can be
$(\epsilon,O(1/\epsilon))$-quantumly tested. Furthermore, the test has one-sided
error.
\end{lemma}
\begin{lemma}
\label{cnlem}
For most $A$ of size $|A| =  n/2$, $P_A$ requires 
$\Omega(\log n)$ queries for a probabilistic $1/3$-test (even for
testers with two-sided error.)
\end{lemma}
%
Before we prove Lemma~\ref{qylem} we note that for any $A$, $P_A$ can
be tested by an (even nonadaptive) one-sided algorithm with 
$O(1/\epsilon + \log n)$; hence, the result of Lemma \ref{cnlem} is tight.
  An $O(1/\epsilon \log n)$-test 
follows from Observation~\ref{ob:gen} below. The slightly more
efficient test is the following: First we query
$x_{2^i}$, $i=0, \ldots ,\log n$. Note that if $x=h(y)$ then $y_i =
x_{2^i}$ for $i=0,\ldots,\log n$. Thus a candidate $y$ for $x=h(y)$ is
found. If $y \notin A$ then $x$ is rejected.  Then for $k =
\operatorname{O} (1 / \epsilon )$ times independently a random index
$i \in \{1,\ldots,n\}$ is chosen and if $x_i \neq y \cdot i$, for the
candidate $y$ that was found before, then $x$ is rejected. Otherwise,
$x$ is accepted. Clearly if $x$ is rejected then $x \notin P_A$. It is
easily verified that if $x$ has Hamming distance more than
$\epsilon n$ from every $z$ in $P_A$ then with constant probability
$x$ is rejected.
\begin{proof}{Proof of Lemma~\ref{qylem}}
$P_A$ can be checked with $\operatorname{O} (1 / \epsilon )$ queries
on a quantum computer: The test is similar to the test above except
that $y$ can be found in $O(1)$ queries: $k$ times query for random $i$, $j$ values
$x_i$, $x_j$, and $x_{i \oplus j}$. If $x_i \oplus x_j \ne x_{i \oplus
  j}$ reject. $k = \operatorname{O} (1 / \epsilon )$ is sufficient to
detect an input $x$ that is $\epsilon n$-far from being a Hadamard
codeword with high probability. Now run the Bernstein-Vazirani
algorithm to obtain $y$. Accept if and only if $y \in A$.  Obviously,
if $x \in P_A$, the given procedure accepts, and if $x$ is far from
any $x' \in P_A$, then it is either far from being a Hadamard codeword
(which is detected in the first step) or it is close to a Hadamard
codeword $h(y')$ for an $y' \not \in A$; note that in this case $x$ is
far from any $h(y)$, $y \in A$ as two distinct Hadamard codewords are
of Hamming distance $n/2$. Thus, in this case the second part
of the tester succeeds with high probability in finding $y'$ and
rejects because $y' \not \in A$. We note also that this algorithm has
one-sided error.
\end{proof}

\begin{proof}{Proof of Lemma~\ref{cnlem}}
The lower bound makes use of the Yao principle \cite{yao:unified}: let
$D$ be an arbitrary probability distribution on positive and negative
inputs (\ie inputs that either belong to $P_A$ or are $\epsilon n$-far
from $P_A$). Then if every deterministic algorithm that makes at most
$q$ queries, errs with probability at least $1/8$ (with respect to input chosen
according to $D$,) then $q$ is a lower bound on the number of queries
of any randomized algorithm for testing $P_A$ with error probability
bounded by $1/8$.

$D$ will be the uniform distribution over Hadamard codewords of length
$n$, namely, generated by choosing $y \in \{0,1\}^{\log n}$ uniformly
at random and setting $x=h(y)$. Note that for any $A \subset
\{0,1\}^{\log n}$, $D$ is concentrated on positive and negative inputs
as required, as two Hadamard codewords are of Hamming distance $n/2$
apart.

The lower bound will be established by a counting argument.  We show
that for a fixed tester that makes $q \leq (\log n)/2$ queries, the
probability (over random choices of $A$) that the the algorithm errs
on at most a 1/8-fraction of the inputs is bounded from above by
$1/(10T)$ where $T$ is the number of such algorithms.  By the union
bound it follows that for most properties there is no such algorithm.

Indeed, let $A \subseteq \{0,1\}^{\log n}$ be chosen by picking
independently each $i \in \{0,1\}^{\log n}$ to be in $A$ with
probability $1/2$; this will not necessarily result in a set $A$ of
size $n/2$ but we can condition on the event that $|A|=n/2$ and will
not loose much.  Let $\mathcal{T}$ be any fixed deterministic decision tree
performing at most $q$ queries in every branch. Then let $c(\mathcal{
  T}) := \{y|\mathcal{ T}(h(y)) = \text{accept} \}$ and let $\mu
(\mathcal{ T}) := |c(\mathcal{T})| /n$, \ie $\mu(\mathcal{T})$ is the
fraction of inputs that $\mathcal{ T}$ accepts.  Assume first that
$\mu(\mathcal{T}) \leq 1/2$. Since for a random $y$ we have $\Pr_y
[\mathcal{ T}(h(y))=\text{accept}]=\mu(\mathcal{T}) \leq 1/2$, it
follows by a Chernoff-type bound that $\Pr_A [|A \cap c(\mathcal{T})|
\geq 3/4 |A|] \leq 2^{-n/8}$. However, if $|A \cap c(\mathcal{T})| <
3/4 |A|$ then $\mathcal{ T}$ will be wrong on at least $1/4$ of the
positive inputs which is at least $n/8$ of all inputs. Hence with
probability at most $2^{-n/8}$, $\mathcal{T}$ will be correct on at
least $7/8$ of the inputs. If $\mu(\mathcal{T}) > 1/2$ the same
reasoning shows that with probability of at most $1-2^{-n/8}$ it will err
on at least an $1/4$-fraction of the negative inputs. Hence in total,
for every fixed $\mathcal{T}$, $\Pr_A[\mathcal{T} \text{ is correct on
  at least } 7/8 \text{ of the inputs}] \leq 2^{-n/8}$.

Now, let us bound from above the number of algorithms that make at
most $q$ queries. As an algorithm may be adaptive, it can be defined
by $2^q-1$ query positions (all queries for all branches) and a
Boolean function $f: \{0, 1\}^q \rightarrow \{ \text{accept},
\text{reject} \}$ of the decision made by the algorithm for the
possible answers. Hence, there are at most $T \le (2n)^{2^q}$ such
algorithms. However, for $q < (\log n)/2$, we have $T \cdot 2^{-n/8} =
\operatorname{o}(1)$, which shows that for most $A$ as above, any
$\epsilon$-test that queries at most $(\log n)/2$ many queries has
error probability of at least $1/8$. Standard amplification techniques
then imply that any algorithm that queries $c \log n$ many queries
(for some constant $c$) has error at least $1/3$.
\end{proof}
 
\begin{observation}\label{ob:gen}
  Let $P \subseteq \{0,1\}^n$ be a property with $|P| =s$ then $P$ can
  be $\epsilon$-tested by a one-sided algorithm using $O((\log
  s)/\epsilon)$ many queries.
\end{observation}
We omit the proof from this draft.

\section{An Exponential Separation}

In this section, we show that a quantum computer can be exponentially
more efficient in testing certain properties than a classical
computer.
\begin{theorem}
  \label{expsepthm}
  There is a languages $L$  that is $(\epsilon,\log n \log \log
  n)$-quantumly testable for any $\epsilon=\Omega (1)$. However, 
  any probabilistic $1/8$-test for $L$ requires $n^{\Omega(1)}$
  queries.
\end{theorem}
The language that we provide is inspired by Simon's problem
\cite{Simon-quantum} and our quantum testing algorithm makes use of
Brassard and H{\o}yer's algorithm for Simon's problem
\cite{brassard&hoyer:simon}. Simon's problem is to find $s \in \{ 0, 1
\}^n \setminus \{ 0^n \} $ from an function-query oracle for some $f :
\{ 0, 1 \}^n \rightarrow \{ 0, 1 \}^n$, such that $f(x) = f(y)
\Leftrightarrow x = y \oplus s$. Simon proved that classically,
$\Omega (2^{n/2})$ queries are required on average to find $s$, and
gave a quantum algorithm for determining $s$ with an expected number
of queries that is polynomial in $n$; Brassard and H{\o}yer improved
the algorithm to worst-case polynomial time. Their algorithm produces
in each run a $z$ with $z \cdot s = 0$ that is linearly independent to
all previously computed such $z$\/s. Essentially, our quantum tester
uses this subroutine to try to extract information about $s$ until it
fails repeatedly.  Recently H{\o}yer~\cite{hoyer:fouriersampling} analyzed this
approach in group-theoretic terms, obtaining an alternative proof to
Theorem~\ref{expqylem}.

In the following, let $N=2^n$ denote the length of the binary string
encoding a function $f : \{ 0, 1 \}^n \rightarrow \{ 0, 1\}$. For $x
\in \{0, 1\}^n$ let $x[j]$ be the $j$\/th bit of $x$, \ie $x = x[1]
\ldots x[n]$; the inner product of $x, y \in \{ 0, 1 \}^n$ as vectors
in $\mathbb{F}_2^n$ is $x \cdot y = \sum_{j=1}^n x[j] y[j] \mod 2$. We
define
\[
L := \{ f \in \{ 0, 1 \}^N : \exists s \in \{ 0, 1 \}^n \setminus \{
0^n \} \; \forall x \in \{ 0, 1 \}^n \; f(x) = f(x \oplus s) \}
\]
Theorem~\ref{expsepthm} follows from the following two theorems.
\begin{theorem}
  \label{expcnlem}
  Any classical $1/8$-tester for $L$ must make $\Omega(\sqrt{N})$
  queries (even when allowing two-sided error.)
\end{theorem}
\begin{theorem}
  \label{expqylem}
  There is a quantum property tester for $L$ making
  $\operatorname{O}(\log N \log \log N)$ queries. Moreover, this
  quantum property tester makes all its queries nonadaptively.
\end{theorem}
\begin{proof}{Proof of Theorem~\ref{expcnlem}}
We again apply the Yao principle \cite{yao:unified} as in the proof of
Lemma~\ref{cnlem}: we construct two distributions, $P$ and
$U$, on positive and negative (at least $N/8$-far) inputs,
respectively, such that any deterministic (adaptive) decision tree
$\mathcal{T}$ has error $1/2-\operatorname{o}(1)$ when trying to
distinguish whether an input is chosen from $U$ or $P$. Indeed, we will show a stronger statement: Let $\mathcal{T}$ be any
deterministic decision tree. Let $v$ be a vertex of $\mathcal{T}$. Let
$\Pr_P(v)$ and $\Pr_U(v)$ be the probability the an input chosen
according to $P$ and $U$, respectively, is consistent with $v$.  We
will show that for any vertex $v$ of $\mathcal{T}$
$|\Pr_P(v)-\Pr_U(v)|= \operatorname{o} (1)$; hence, $\mathcal{T}$ has
error $1/2-\operatorname{o}(1)$.

The distribution $P$ is defined as follows: We first chose $s \in
\{0,1\}^n$ at random. This defines a matching $M_s$ of $\{0,1\}^n$
by matching $x$ with $x \oplus s$. Now a function $f_s$ is defined by
choosing for each matched pair independently $f_s(x)=f_s(x \oplus
s)=1$ with probability $1/2$ and $f_s(x)=f_s(x \oplus s)=0$ with probability $1/2$. Clearly,
this defines a distribution that is concentrated on positive inputs.
Note that it might be that by choosing different $s$'s we end up
choosing the same function, however, this will be considered different
events in the probability space. Namely, the atomic events in $P$
really are the pairs $(s,f_s)$ as described above.

Now let $U$ be the uniform distribution over all functions, namely, we
select the function by choosing for each $x$ independently $f(x)=1$
with probability $1/2$ and $0$ with probability $1/2$. Since every
function has a non zero probability, $U$ is not supported exclusively
on the negative instances. However, as we proceed to show, a function
chosen according to $U$ is $N/8$-far from having the property with
very high probability, and hence $U$ will be a good approximation to
the desired distribution:
\begin{definition}
  For $f: \{0,1\}^n \rightarrow \{0,1\}$ and $s \in \{0,1\}^n$ we
  define $n_s := |\{x: f(x)=f(x \oplus s) \} |$.
\end{definition}
\begin{lemma}\label{cl:lb2}
  Let $f$ be chosen according to $U$. Then $\Pr_U[\exists s\in
  \{0,1\}^n : n_s \geq N/8] \leq \exp(-\Omega(N))$.
\end{lemma}
\begin{proof}{Proof}
  Let $f$ be chosen according to $U$ and $s \in \{ 0, 1 \}^n$. By a
  Chernoff bound we obtain $\Pr_U[n_s \geq N/8] \leq
  \exp(-\Omega(N))$. Together with the union bound over all $s$'s this
  yields $\Pr_U[\exists s\in \{0,1\}^n : n_s \geq N/8] \leq 2^n \cdot
  \exp(-\Omega(N)) \leq \exp(-\Omega(N))$.
\end{proof}
In particular, a direct consequence of Lemma~\ref{cl:lb2} is that with
probability $1-\exp(-\Omega(N))$ an input chosen according to $U$ will
be $N/8$-far from having the property. 

From the definition of $U$, we immediately obtain the following:
\begin{lemma}\label{cl:lb5}
  Let $\mathcal{T}$ be any fixed deterministic decision tree and let $v$ be a
  vertex of depth $d$ in $\mathcal{T}$. Then \\ $\Pr_U[f \text{ is
    consistent with the path to } v]=2^{-d}$.
\end{lemma}
We now want to derive a similar bound as in the last lemma for
functions chosen according to $P$. For this we need the following
definition for the event that after $d$ queries, nothing has been
learned about the hidden $s$:
\begin{definition}
  Let $\mathcal{T}$ be a deterministic decision tree and $u$ a vertex in
  $\mathcal{T}$ at depth $d$. We denote the path from the root of
  $\mathcal{T}$ to $u$ by $\operatorname{path}(u)$. Any vertex $v$ in
  $\mathcal{T}$ defines a query position $x_v \in \{0,1\}^n$. For
  $f=f_s$ chosen according to $P$, we denote by $B_u$ the event $B_u :=
  \{ (s,f_s) : s \neq x_{v} \oplus x_{w} \text{ for all } v,w \in \operatorname{path}(u)\}$.
\end{definition}
\begin{lemma}\label{cl:lb6}
  Let $v$ be a vertex of depth $d$ in a decision tree $\mathcal{T}$.
  Then $\Pr_P[B_v] \geq 1 - \binom{d-1}{2}/N$
\end{lemma}
\begin{proof}{Proof}
  $B_v$ does not occur if for some $v,w$ on the path to $v$ we have
  $s=x_v \oplus x_w$. As there are $d-1$ such vertices, there are at
  most $\binom{d-1}{2}$ pairs.  Each of these pair excludes exactly one
  $s$ and there are $N$ possible $s$'s.
\end{proof}
\begin{lemma}\label{cl:lb7}
  Let $v$ be a vertex of depth $d$ in a decision tree $\mathcal{T}$
  and let $f$ be chosen according to $P$. Then $\Pr_P[f \text{ is
    consistent with } v|B_v] = 2^{-d}$.
\end{lemma}
\begin{proof}{Proof}
  By the definition of $P$, $f$ gets independently random values on
  vertices that are not matched. But if $B_v$ occurs, then no two
  vertices along the path to $v$ are matched and hence the claim
  follows.
\end{proof}
Now we can complete the proof of the theorem: assume that $\mathcal{T}$ is
a deterministic decision tree of depth $d=\operatorname{o}(\sqrt{N})$
and let $v$ be any leaf of $\mathcal{T}$. Then by Lemmas \ref{cl:lb6}
and \ref{cl:lb7}, we get that $\Pr_P[f \text{ is consistent with } v]
= (1-\operatorname{o}(1))2^{-d}$.  On the other hand, let $U'$ be the
distribution on negative inputs defined by $U$ conditioned on the
event that the input is at least $N/8$-far from the property.  Then by
Lemmas \ref{cl:lb2} and \ref{cl:lb5} we get that $\Pr_{U'}[f \text{ is
  consistent with } v] = (1-\operatorname{o}(1))2^{-d}$ and hence
$\mathcal{T}$ has only $\operatorname{o}(1)$ bias of being right on
every leaf which directly imply that its error probability is $1/2 -
\operatorname{o}(1)$.
\end{proof}

\begin{proof}{Proof of Theorem~\ref{expqylem}}
  We give a quantum algorithm making $\operatorname{O}(\log N \log
  \log N)$ queries to the quantum oracle for input $f \in \{0, 1\}^N$.
  We will show that it accepts with  probability 1 if $f \in L$ and rejects with
  high probability if the Hamming distance between $f$ and every $g
  \in L$ is at least $\epsilon N$. Our algorithm, given in
  Figure~\ref{fig:expalg}, consists of a classical main program
  \begin{figure}[htbp]
    \centering
    \begin{tabular}{@{}p{0.5\textwidth}@{}p{0.5\textwidth}@{}}
      \begin{algorithmic}
        \STATE \textsc{Main Program}
        \FOR{$k=0$ to $n-1$}
        \STATE $l \leftarrow 0$
        \REPEAT
        \STATE $z \leftarrow Q(z_{1}, \ldots, z_{k})$
        \STATE $l \leftarrow l + 1$
        \UNTIL{$z \ne 0$ or $l>2(\log n)/\epsilon^2$}
        \IF{$z=0$}
        \STATE accept
        \ELSE
        \STATE $z_{k+1} \leftarrow z$
        \ENDIF
        \ENDFOR
        \STATE reject
      \end{algorithmic}
      &
      \begin{algorithmic}
        \STATE \textsc{Subroutine $Q$}
        \STATE {\bf input:} $z_{1}, \ldots, z_{k} \in \{ 0, 1 \}^n$
        \STATE {\bf output:} $z \in \{ 0, 1 \}^n$
        \STATE {\bf quantum workspace:} $\mathcal X \tensor \mathcal Y
        \tensor \mathcal Z$
        \STATE (where $\mathcal X$ is $n$ qubits $\mathcal X = \mathcal X_1 \tensor \cdots \tensor \mathcal X_n$, $\mathcal X_i = \mathbb C^2$,
        \STATE $\mathcal Y = \mathbb C^2$ is one qubit, and
        \STATE $\mathcal Z$ is $k$ qubits $\mathcal Z = \mathcal Z_1 \tensor \cdots \tensor \mathcal Z_k$, $\mathcal Z_j = \mathbb C^2$)
        \STATE initialize the workspace to 
        $\ket{0^n} \ket 0 \ket{0^k}$
        \STATE apply $H_{2^n}$ to $\mathcal X$
        \STATE apply $O_f$ to $\mathcal X \tensor \mathcal Y$
        \STATE apply $H_{2^n}$ to $\mathcal X$
        \FOR{$j=1$ to $k$}
        \STATE $i \leftarrow \min \{i : z_{j}[i] = 1 \}$
        \STATE apply $\operatorname{CNOT}$ with control $\mathcal X_i$ and target $\mathcal Z_j$
        \STATE apply $\ket x \mapsto \ket{x \oplus z_{j}}$ to $\mathcal X$ conditional on $\mathcal Z_j$
        \STATE apply $H_2$ to $\mathcal Z_j$
        \ENDFOR
        \STATE {\bf return} measurement of $\mathcal X$
      \end{algorithmic}
    \end{tabular}
    \caption{The Quantum Property Tester}
    \label{fig:expalg}
  \end{figure}
  and a quantum subroutine $Q$ adapted from Brassard and H{\o}yer's
  algorithm for Simon's problem \cite[Section
  4]{brassard&hoyer:simon}. The quantum gates used are the
  $2^n$-dimensional Hadamard transform $H_{2^n}$, which applies $
  \begin{pmatrix}
    1 & 1 \\ 1 & -1
  \end{pmatrix} / \sqrt{2}$ individually to each of $n$ qubits, the 
  quantum oracle query $O_f$, and classical reversible operations run
  in quantum superposition.
  
  The following technical lemma captures the operation of the quantum
  subroutine $Q$. Essential proofs to this and later lemmas are
  included in the appendix.
  \begin{lemma}
    \label{expstate}
    When $Q$ is passed $k$ linearly independent vectors $z_{1},
    \ldots, z_{k}$ so that all $i_j := \min \{i : z_{j}[i] = 1 \}$
    are distinct for $1 \le j \le k$, then the state before the
    measurement is
    \[
    \ket \psi := \frac{\sqrt{2^{k}}}{N} \sum_{x \in \{ 0, 1 \}^n}
    \sum_{\substack{ y \in \{ 0, 1 \}^n \\ y[i_j]=0 \: \forall j \le
        k }} (-1)^{x \cdot y} \ket{y} \ket{f(x)} \ket{x \cdot z_{1}}
    \cdots \ket{x \cdot z_{k} } .
    \]
  \end{lemma}
  As an immediate consequence, we can establish the invariant that in
  the main program $\{ z_{1}, \ldots, z_{k} \}$ always is linearly
  independent with $i_j = \min \{i : z_{j}[i] = 1 \}$ distinct for $1
  \le j \le k$; moreover, if $f \in L$, then just as in Simon's
  algorithm, a nonzero $z$ is orthogonal to the hidden $s$:
  \begin{lemma}
    \label{expz}
    If measuring the first register, $\mathcal X$, yields a nonzero
    value $z$, then
    \begin{enumerate}
    \item $\{ z_{1}, \ldots, z_{k}, z \}$ is linearly
      independent,
    \item $\min \{ i : z[i] = 1 \}$ is distinct from $i_j$ for $1 \le
      j \le k$, and
    \item if $f \in L$, then $z \cdot s = 0$ for any $s$ such that
      $f(x) = f(x \oplus s)$ for all $x$.
    \end{enumerate}
  \end{lemma}
  Next, we want to assess the probability of obtaining $z=0$ in the
  main loop. We let $P_0$ denote the projection operator mapping $\ket
  0 \ket y \ket z \mapsto \ket 0 \ket y \ket z$ and $\ket x \ket y
  \ket z \mapsto 0$ for $x \ne 0$; hence, $\norm{P_0 \ket \psi}^2$ is
  the probability of obtaining $0$ when measuring subspace $\mathcal
  X$ of the quantum register in state $\ket \psi$.
  We can characterize the probability for outcome $z=0$ in terms of
  the following definition and lemma:
  \begin{definition}
    For $c \in \{ 0, 1 \}^k$ and $z_1$, \ldots, $z_k \in \{ 0, 1 \}^n$
    we define $D_c := \{ x \in \{ 0, 1 \}^n : x \cdot z_{1} = c[1],
    \ldots, x \cdot z_{k} = c[k] \}$.
  \end{definition}
  \begin{lemma}
    \label{expzerostate}
    Let $\ket \psi$ be the state before the measurement in $Q$, when
    $Q$ is passed $k$ linearly independent vectors $z_{1}, \ldots,
    z_{k}$ so that all $i_j := \min \{i : z_{j}[i] = 1 \}$ are
    distinct for $1 \le j \le k$.
    \begin{enumerate}
    \item\label{expzerostateexact} $\norm{P_0 \ket \psi}^2 = 1$ if and
      only if for every $c \in \{ 0, 1\}^k$, $f$ is constant when
      restricted to $D_c$.
    \item\label{expzerostateapprox} If $\norm{P_0 \ket \psi}^2 \ge 1 -
      \epsilon^2/2$, then $f$ differs in at most $\epsilon N$ points
      from some function $g$ that is constant when restricted to $D_c$
      for every $c \in \{ 0, 1 \}^k$.
    \end{enumerate}
  \end{lemma}
  We need to relate these two cases to membership in $L$ and bound the
  number of repetitions needed to distinguish between the two cases.
  This is achieved by the following two lemmas.
  \begin{lemma}
    \label{expmember}
    Let $k$ be the minimum number of linearly independent vectors
    $z_{1}$, \ldots, $z_{k}$ so that for each $c \in \{ 0 , 1 \}^k$,
    $f$ is constant when restricted to $D_c$. Then $f \in L$ if and
    only if $k<n$.
  \end{lemma}
  \begin{lemma}
    \label{expmeasurezero}
    Let $0 < q < 1$, and $\ket{\phi_1}$, \ldots, $\ket{\phi_m}$ be
    quantum states satisfying $\norm{P_0 \ket{\phi_j} }^2 < 1 -
    \delta$ for $1 \le j \le m$. If $m=\log q / \log (1-\delta) =
    \Theta(- \log q/\delta)$, then with probability at most $q$
    measuring the $\mathcal X$ register of $\ket{\phi_1}$, \ldots,
    $\ket{\phi_m}$ will yield $m$ times outcome 0.
  \end{lemma}
  \begin{proof}{Proof}
    $
    \Pr \left[ m \text{ times } 0 \left| \forall j: \norm{P_0
          \ket{\phi_j} }^2 < 1 - \delta \right. \right] < ( 1 -
    \delta)^m = (1-\delta)^{\log q / \log (1-\delta)} = q
    $.
  \end{proof}
  Now all the ingredients for wrapping up the argument are at hand;
  first consider $f \in L$.  Let $S := \{ s : f(x) = f(x \oplus s) \;
  \forall x \}$ be the set of all ``Simon promises'' of $f$, and
  $S^\perp := \{ z : z \cdot s = 0 \; \forall s \in S \}$ the vectors
  that are orthogonal to all such promises. By Lemma~\ref{expz} the
  nonzero $z$ computed by the algorithm lie in $S^\perp$ and are
  linearly independent, therefore after $\dim S^\perp$ rounds of the
  main loop, we measure $z=0$ with certainty. Since $f \in L$, $\dim
  S>0$ and thus $\dim S^\perp < n$.
  
  If $f$ is $\epsilon n$-far from being in $L$, then by
  Lemma~\ref{expmember} $f$ is $\epsilon n$-far from being close to a
  function for which a $k<n$ and $z_{1}$, \ldots, $z_{k}$ exist so
  that $f$ is constant when restricted to $D_c$ for any of the $c \in
  \{ 0, 1 \}^k$. Therefore, by Lemma~\ref{expzerostate}
  case~\ref{expzerostateapprox}, for all $k<n$, $\norm{P_0 \ket
    \psi}^2 < 1 - \epsilon^2/2$. Thus, Lemma~\ref{expmeasurezero}
  guarantees that we accept with probability at most $1/3$ if we let
  $q=1/(3n)$ and thus $m \le 2 (\log n)/\epsilon^2$.
\end{proof}

\section{Quantum Lower Bounds}

In this section we prove that not every language has a quantum
property tester.
\begin{theorem}
  Most properties containing $2^{n/20}$ elements of\/ $\{0,1\}^n$
  require quantum property testers using $\Omega(n)$ queries.
\end{theorem}
\begin{proof}{Proof}
  Fix $n$, a small $\epsilon$, and a quantum algorithm $A$ making $q
  := n/400$ queries. Pick a property $P$ as a random subset of
  $\{0,1\}^n$ of size $2^{n/20}$.  Let $P_\epsilon := \{ y : d(x,y) <
  \epsilon n \text{ for some } x \in P \}$; using
  $\sum_{k=0}^{\epsilon n} \binom{n}{k} \le 2^{H(\epsilon) n}$ where
  $H(\epsilon) = - \epsilon \log \epsilon - (1-\epsilon) \log
  (1-\epsilon)$, we obtain $|P_\epsilon| \le 2^{(1/20+H(\epsilon))
    n}$. In order for $A$ to test properties of size $2^{n/20}$, it
  needs to reject with high probability on at least
  $2^n-2^{(1/20+H(\epsilon))n}$ inputs; but then, the probability that
  $A$ accepts with high probability on a random $x \in \{ 0, 1 \}^n$
  is bounded by $2^{(1/20+H(\epsilon))n}/2^n$ and therefore the
  probability that $A$ accepts with high probability on $|P|$ random
  inputs is bounded by $2^{-(1-1/20-H(\epsilon)) n |P|} = 2^{-2^{ n/20
      + \Theta(\log n) }}$.
  
  We would like to sum this success probability over all algorithms
  using the union bound to argue that for most properties no algorithm
  can succeed. However, there is an uncountable number of possible
  quantum algorithms with arbitrary quantum transitions. But by Beals,
  Buhrman, Cleve, Mosca, and de~Wolf~\cite{BBCMW}, the acceptance
  probability of $A$ can be written as a multilinear polynomial of
  degree at most $2q$ where the $n$ variables are the bits of the
  input; using results of Bennett, Bernstein, Brassard, and
  Vazirani~\cite{BBBV} and Solovay and Yao~\cite{SoYa}, any quantum
  algorithm can be approximated by another algorithm such that the
  coefficients of the polynomials describing the accepting probability
  are integers of absolute value less than
  $2^{n^{\operatorname{O}(1)}}$ over some fixed denominator. There are
  less than $2^{n H(2q/n)}$ degree $2q$ polynomials in $n$ variables,
  thus we can limit ourselves to $2^{n^{\operatorname{O}(1)} 2^{n
      H(2q/n)}} \le 2^{2^{n/20 \cdot 91/100 + \Theta ( \log n)}}$ algorithms.
  
  Thus, by the union bound, for most properties of size $2^{n/20}$, no
  quantum algorithm with $q$ queries will be a tester for it.
\end{proof}

We also give an explicit natural property that requires a large number
of quantum queries to test.
\begin{theorem}
  \label{lbthm}
  The range of a $d$-wise independent pseudorandom generator requires
  $(d+1)/2$ quantum queries to test for any odd $d \le n/\log n - 1$.
\end{theorem}
We will make use of the following lemma:
\begin{lemma}[see~\cite{ABI}]
\label{indlem}
Suppose $n=2^k-1$ and $d= 2 t + 1 \le n$. Then there exists a uniform
probability space $\Omega$ of size $2 ( n + 1)^t$ and $d$-wise
independent random variables $\xi_1$, \ldots, $\xi_n$ over $\Omega$
each of which takes the values 0 and 1 with probability $1/2$.
\end{lemma}
The proof of Lemma~\ref{indlem} is constructive and the construction
uniform in $n$; for given $n$ and $d$, consider the language $P$ of
bit strings $\xi ( z ) := \xi_1 ( z ) \ldots \xi_n ( z
)$ for all events $z \in \Omega = \{ 1, \ldots, 2 ( n + 1)^t \}
$.  Classically, deciding membership in $P$ takes more than $d$
queries: for all $d$ positions $i_1$, \ldots, $i_d$ and all strings
$v_1 \ldots v_d \in \{ 0, 1 \}^d$ there is an $z$ such that
$\xi_{i_1} ( z ) \ldots \xi_{i_d} ( z ) = v_1 \ldots v_d$.
On the other hand, $\lfloor \log | \Omega | \rfloor + 1 =
\operatorname{O} ( d \log n)$ queries are always sufficient.
\begin{proof}{Proof of Theorem~\ref{lbthm}}
  A quantum computer deciding membership for $x \in \{ 0, 1 \}^n$ in
  $P := \{ \xi ( z ) : z \in \Omega \}$ with $T$ queries
  gives rise to a degree $2 T$ approximating (multilinear
  $n$-variable) polynomial $p(x)=p(x_1, \ldots, x_n)$ (see
  \cite{BBCMW}.)  We show that there must be high-degree monomials in
  $p$ by comparing the expectation of $p(x)$ for randomly chosen $x
  \in \{ 0, 1 \}^n$ with the expectation of $p(x)$ for randomly chosen
  $x \in P$.
  
  For uniformly distributed $x \in \{ 0, 1 \}^n$, we have $\E [ p(x) |
  x \in P ] \ge 2/3$ and $\E [ p(x) | x \notin P ] \le 1/3$. Since of
  $|P|=\operatorname{o}(2^n)$, $\E [ p(x) ] \le 1/3 +
  \operatorname{o}(1)$ and thus $\Delta := \E [ p(x) | x \in P ] - \E
  [ p(x) ] \ge 1/3 - \operatorname{o} (1)$.  Considering $p(x) =
  \sum_i \alpha_i m_i (x)$ as a linear combination of $n$-variable
  multilinear monomials $m_i$, we have by the linearity of expectation
  $ \E [ p(x_1, \ldots, x_n) ] = \sum_i \alpha_i \E [ m_i(x_1, \ldots,
  x_n) ] $. Because of the $d$-wise independence of the bits of any $x
  \in P$, for any $m_i$ of degree at most $d$ holds $ \E [ m_i(x) ] =
  \E [ m_i(x) | x \in P ] $. Since $\Delta > 0$, $p$ must comprise
  monomials of degree greater than $d$.  Hence, the number of queries
  $T$ is greater than $d/2$.
  
  This proof extends in a straightforward manner to the case of
  testing the property $P$: let again $P_\epsilon := \{ y : d(x,y) <
  \epsilon n \text{ for some } x \in P \}$. Then $|P_\epsilon| \le
  2^{H(\epsilon) n} |P| = \operatorname{O} (2^{H(\epsilon) n + d \log
    n})$, so $\E [ p(x) ] = |P_\epsilon|/2^n \E [ p(x) | x \in
  P_\epsilon ] + (1-|P_\epsilon|)/2^n \E [ p(x) | x \notin P_\epsilon
  ] \le 1/3 + \operatorname{o}(1)$ for any $d = n/\log n -
  \omega(1/\log n)$ and any $\epsilon$ with $H(\epsilon) = 1 -
  \omega(1/n)$.
\end{proof}


\section{Further Research}

Our paper opens the door to the world of quantum property testing.
Several interesting problems remain including
\begin{itemize}
\item Can one get the greatest possible separation of quantum and
  classical property testing, \ie is there a language that requires
  $\Omega(n)$ classical queries but only $\operatorname{O}(1)$ quantum
  queries to test?
\item Are there other natural problems that do not have quantum
  property testers? We conjecture for instance that the language
  $\{uuvv : u,v \in \Sigma^* \}$ does not have a quantum property
  tester.
\item Beals, et.\ al.~\cite{BBCMW} observed that any $k$-query quantum
  algorithm gives rise to a degree-$2k$ polynomial in the input bits,
  which gives the acceptance probability of the algorithm; thus, a
  quantum property tester for $P$ gives rise to a polynomial that is
  on all binary inputs between $0$ and $1$, that is at least $2/3$ on
  inputs with the property $P$ and at most $1/3$ on inputs far from
  having the property $P$. Szegedy~\cite{szegedy:testingpolyconj}
  suggested to algebraically characterize the complexity of classical
  testing by the minimum degree of such polynomials; as mentioned in
  the introduction, our results imply that this cannot be the case for
  classical testers. However, it is an open question whether quantum
  property testing can be algebraically characterized in this way.
\end{itemize}
We hope that further research will lead to a greater understanding of
what can and cannot be tested with quantum property testers.

\section*{Acknowledgments}

We thank Ronitt Rubinfeld for discussions and pointers on property
testing.


\newcommand{\etalchar}[1]{$^{#1}$}



\appendix

\section{Appendix}

\begin{proof}{Proof of Lemma~\ref{expstate}}
  We follow the steps of subroutine $Q$ when passed $k$ linearly
  independent vectors $z_{1}, \ldots, z_{k}$ so that all $i_j :=
  \min \{i : z_{j}[i] = 1 \}$ are distinct for $1 \le j \le k$.
  \[
  \ket{0^n} \ket 0 \ket{0^k} \longmapsto \frac{1}{\sqrt{N}} \sum_{x \in
    \{ 0, 1 \}^n} \ket{x} \ket 0 \ket{0^k} \longmapsto
  \frac{1}{\sqrt{N}} \sum_{x \in \{ 0, 1 \}^n} \ket{x} \ket{f(x)}
  \ket{0^k} \longmapsto \frac{1}{N} \sum_{x, y \in \{ 0, 1 \}^n} (-1)^{x
    \cdot y} \ket{y} \ket{f(x)} \ket{0^k}
  \]
  This is the state before the \textbf{for} loop is entered. We claim
  (and proceed to show by induction) that after the $J$\/th execution of
  the loop body, the state is
  \[
  \frac{\sqrt{2^{J}}}{N} \sum_{x \in \{ 0, 1 \}^n} \sum_{\substack{ y
      \in \{ 0, 1 \}^n \\ y[i_j]=0 \: \forall j \le J }} (-1)^{x \cdot
    y} \ket{y} \ket{f(x)} \ket{x \cdot z_{1}} \cdots \ket{x \cdot
    z_{J}} \ket{0^{k-J}} .
  \]
  Executing the body of the loop for $j=J+1$,
  \begin{gather*}
    \frac{\sqrt{2^{J}}}{N} \sum_{x \in \{ 0, 1 \}^n} \sum_{\substack{
        y \in \{ 0, 1 \}^n \\ y[i_j]=0 \: \forall j \le J }} (-1)^{x
      \cdot y} \ket{y} \ket{f(x)} \ket{x \cdot z_{1}} \cdots \ket{x
      \cdot z_{J}} \ket{0} \ket{0^{k-J-1}} \longmapsto \\
    \frac{\sqrt{2^{J}}}{N} \sum_{x \in \{ 0, 1 \}^n} \sum_{\substack{
        y \in \{ 0, 1 \}^n \\ y[i_j]=0 \: \forall j \le J }} (-1)^{x
      \cdot y} \ket{y} \ket{f(x)} \ket{x \cdot z_{1}} \cdots \ket{x
      \cdot z_{J}} \ket{y[i_{j+1}]} \ket{0^{k-J-1}} = \\
    \frac{\sqrt{2^{J}}}{N} \sum_{x \in \{ 0, 1 \}^n} \sum_{b\in
      \{0,1\} } \sum_{\substack{ y \in \{ 0, 1 \}^n \\ y[i_j]=0 \:
        \forall j \le J+1 }} (-1)^{x \cdot (y \oplus b z_{J+1} )}
    \ket{y \oplus b z_{J+1} } \ket{f(x)} \ket{x \cdot z_{1}}
    \cdots \ket{x \cdot
      z_{J}} \ket{b} \ket{0^{k-J-1}} \longmapsto \\
    \frac{\sqrt{2^{J}}}{N} \sum_{x \in \{ 0, 1 \}^n} \sum_{b\in
      \{0,1\} } \sum_{\substack{ y \in \{ 0, 1 \}^n \\ y[i_j]=0 \:
        \forall j \le J+1 }} (-1)^{x \cdot (y \oplus b z_{J+1} )}
    \ket{y} \ket{f(x)} \ket{x \cdot z_{1}} \cdots \ket{x \cdot
      z_{J}} \ket{b}
    \ket{0^{k-J-1}} = \\
    \frac{\sqrt{2^{J+1}}}{N} \sum_{x \in \{ 0, 1 \}^n}
    \sum_{\substack{ y \in \{ 0, 1 \}^n \\ y[i_j]=0 \: \forall j \le
        J+1 }} (-1)^{x \cdot y} \ket{y} \ket{f(x)} \ket{x \cdot
      z_{1}} \cdots \ket{x \cdot z_{J}} \frac{1}{\sqrt 2}
    \sum_{b\in \{0,1\} } (-1)^{x
      \cdot ( b z_{J+1} )} \ket{b} \ket{0^{k-J-1}} \longmapsto \\
    \frac{\sqrt{2^{J+1}}}{N} \sum_{x \in \{ 0, 1 \}^n}
    \sum_{\substack{ y \in \{ 0, 1 \}^n \\ y[i_j]=0 \: \forall j \le
        J+1 }} (-1)^{x \cdot y} \ket{y} \ket{f(x)} \ket{x \cdot
      z_{1}} \cdots \ket{x \cdot z_{J+1}} \ket{0^{k-J-1}}
  \end{gather*}
\end{proof}
\begin{proof}{Proof of Lemma~\ref{expzerostate}}
  For $b \in \{ 0, 1\}$ let $D_{b,c} := D_c \cap f^{-1} \{ b \} = \{ x
  : f(x)=b \text{ and } x \cdot z_{1} = c[1], \ldots, x \cdot z_{k} =
  c[k] \}$.  Note that the $D_{b,c}$ and $D_c$ also depend on $z_{1}$,
  \ldots, $z_{k}$ and the $D_{b,c}$ depend on $f$.  Let
  \[
  \ket{\psi_0} := \frac{\sqrt{2^{k}}}{N} \sum_{x \in \{ 0, 1 \}^n}
  \ket{0} \ket{f(x)} \ket{x \cdot z_{1}} \cdots \ket{x \cdot z_{k} } =
  \frac{\sqrt{2^{k}}}{N} \sum_{b \in \{ 0, 1 \}} \sum_{c \in \{ 0, 1
    \}^k } | D_{b,c} | \ket{0} \ket{b} \ket{c[1]} \cdots \ket{c[k]} .
  \]
  By Lemma~\ref{expstate}, at the end of $Q$, the system is in state
  \[
  \ket \psi = \ket{\psi_0} + \frac{\sqrt{2^{k}}}{N} \sum_{x \in \{ 0,
    1 \}^n} \; \sum_{\substack{ y \in \{ 0, 1 \}^n \setminus \{ 0 \} \\
      y[i_j]=0 \: \forall j \le k }} (-1)^{x \cdot y} \ket{y}
  \ket{f(x)} \ket{x \cdot z_{1}} \cdots \ket{x \cdot z_{k} } .
  \]
  We consider the case $\norm{P_0 \ket \psi}^2=1$. Then the register
  $\mathcal X$ must be in state $\ket 0$ and thus $\ket \psi =
  \ket{\psi_0}$. Since the state has norm 1, we know that
  \begin{equation}
    \label{eq:expsqr}
    \sum_{b \in \{ 0, 1\}} \sum_{c \in \{ 0, 1 \}^k } | D_{b,c} |^2 =
    \frac{N^2}{2^{k}}  
    .
  \end{equation}
  The $D_{b,c}$ partition $\{ 0, 1\}^n$ and the $D_c = D_{0, c} \cup
  D_{1, c}$ have the same size for all $c \in \{ 0, 1\}^k$ (because
  they are cosets of $D_0$.) Therefore,
  \begin{equation}
    \label{eq:explinconstr}
    \sum_{b \in \{0, 1\}} \sum_{c \in \{ 0, 1 \}^k} | D_{b,c} | = N
    \qquad \text{and} \qquad |D_{0,c}| +
    |D_{1,c}| = \frac{N}{2^k} \text{ for all } c \in \{ 0, 1 \}^k.
  \end{equation}
  $|D_{0,c}|^2 + |D_{1,c}|^2 \le N^2/{2^{2k}}$, but in order for
  equation~(\ref{eq:expsqr}) to hold, $|D_{0,c}|^2 + |D_{1,c}|^2$ must
  be exactly $N^2/{2^{2k}}$. This can only be achieved if either
  $D_{0,c}$ or $D_{1,c}$ is empty. So $f$ must be constant when
  restricted to $D_c$ for any $c \in \{ 0, 1\}^k$.  Conversely, if $f$
  is constant when restricted to $D_c$ for any $c \in \{ 0, 1\}^k$,
  then equation~(\ref{eq:expsqr}) holds, therefore
  $\norm{\ket{\psi_0}} = 1$ and $\ket{\psi} = \ket{\psi_0}$.  This
  concludes the proof of case~\ref{expzerostateexact} of the lemma.
  
  If $\norm{P_0 \ket \psi}^2 = \norm{\ket{\psi_0}}^2 \ge 1 - \delta$,
  then
  \begin{equation}
    \label{eq:expsqrapprox}
    \sum_{b \in \{ 0, 1\}} \sum_{c \in \{ 0, 1 \}^k } | D_{b,c} |^2 \ge 
    \left(1 - \delta \right) \frac{N^2}{2^{k}}
  \end{equation}
  Still, the constraints~(\ref{eq:explinconstr}) hold; let $r2^k$ be
  the number of $c \in \{ 0, 1 \}^k$ so that $\min\{ |D_{0,c}|,
  |D_{1,c}| \} \ge \gamma N/2^k$. Then
  \[
  \sum_{b \in \{ 0, 1\}} \sum_{c \in \{ 0, 1 \}^k } | D_{b,c} |^2 \le
  r 2^k (\gamma^2 + (1-\gamma)^2) \frac{N^2}{2^{2k}} + (1 - r) 2^k
  \frac{N^2}{2^{2k}},
  \]
  and using (\ref{eq:expsqrapprox}), we obtain $r \le
  \delta/(1-\gamma^2-(1-\gamma)^2)$. With $\delta = \epsilon^2 / 2$
  and $\gamma = \epsilon/2$, this implies $r \le \epsilon$. But then
  \[
  \sum_{c \in \{ 0, 1 \}^k } \min \left\{ |D_{0,c}|, |D_{1,c}|
  \right\} \le r2^k \frac{N}{2^{k+1}} + (1-r) 2^k \gamma \frac{N}{2^k}
  \le \epsilon N
  \]
\end{proof}
\begin{proof}{Proof of Lemma~\ref{expmember}}
  If $k<n$, then there exists an $s$ with $s \cdot z_{1}=0, \ldots, s
  \cdot z_{k} = 0$. For any such $s$ and any $x$, we have $x \cdot
  z_{1} = (x \oplus s) \cdot z_{1}, \ldots, x \cdot z_{k} = (x \oplus
  s) \cdot z_{k}$ and $x \in D_{f(x), x \cdot z_{1}, \ldots, x \cdot
    z_{k}}$ and $x \oplus s \in D_{f(x \oplus s), x \cdot z_{1},
    \ldots, x \cdot z_{k}}$, therefore $f(x)=f(x \oplus s)$.
  Conversely, for $f \in L$, $S := \{ s : \forall x f(x) = f(x \oplus
  s) \}$ is a non-trivial subspace of $\{ 0, 1 \}^n$, therefore
  $S^\perp = \{ z : z \cdot s = 0 \forall s \in S \}$ is a proper
  subspace of $\{ 0, 1 \}^n$. Let $z_{1}$, \ldots, $z_{k}$ be any
  basis of $S^\perp$.
\end{proof}

\end{document}